\documentclass[11pt]{article}
\usepackage{iap2000,epsfig}

\def\be{\begin{equation}}
\def\ee{\end{equation}}
\def\bea{\begin{eqnarray}}
\def\eea{\end{eqnarray}}

\begin{document}
\vspace*{4cm}
\title{
THE LOW SURFACE BRIGHTNESS MEMBERS OF NEARBY GROUPS}

\author{C. Mendes de Oliveira, Eleazar Rodrigo Carrasco Damele}

\address{Departamento de Astronomia, Instituto Astron\^omico e Geof\'{\i}sico, 
Universidade de S\~ao Paulo, Brazil}

\author{Leopoldo Infante}
\address{Departamento de Astronom\'{\i}a y Astrof\'{\i}sica, 
Facultad de F\'{\i}sica, 
Pontificia Universidad Cat\'olica de Chile}
 
\author{Michael Bolte}
\address{Lick Observatory,
       Board of Studies in Astronomy and Astrophysics,  
University of
       California, Santa Cruz, USA}

\maketitle

\abstracts{
   We are undertaking a large, wide-field survey of nearby groups and
poor clusters. The main goals are identifying and characterizing the
properties of low-surface-brightness dwarf galaxies and determining
the galaxy luminosity function for M$_R > -17$. Large areas (typically
0.3--0.5 degree$^2$ per system but up to 7 degree$^2$) of the groups
Telescopium, Leo I, Dorado, N5575, HCG 42, HCG 44,  HCG 68 and the
poor clusters IC 4765 and Hydra have so far been surveyed in V and I.
We present the preliminary results for the photometric study of the
groups HCG 42, Telescopium (or NGC6868) and IC4765.  Hundreds of new low
surface brightness galaxies are catalogued. Their spatial distributions,
colors, types and sizes will be studied as a function of the richness
of their environments.}

\section{Introduction}

Dwarf galaxies are the most common type of galaxies in the local
universe. In addition, they are thought to be the single systems with
the largest dark-matter content, with M/L ratios as high as that of
groups and poor clusters \cite{car88}. Hence their spatial distribution
and mass spectrum may give us important insight into the spatial scales
over which mass is distributed in the universe.  However, they are also
the hardest galaxies to observe, due to their low-surface brightnesses
and low luminosities.  Only recently it has become possible to obtain
deep, wide-field CCD {\it photometry} of significant numbers of these
galaxies, mostly in clusters. {\it Spectroscopy} of such low luminosity
systems is, however, still a challenge even with large-class
telescopes unless emission lines are present.

Dwarf galaxies in nearby clusters and groups have been studied in detail
with photographic 
material \cite{tho93,san85,imp88,fer89,bot91,fer91}.
 Only recently, with the advent of
large-format CCDs, have these searches become possible in wide-field
CCD images of nearby groups and clusters \cite{one97,jer00}.
From these works we learned that the dwarf elliptical (dE/dS0) galaxies
dominate the faint-end of the number counts in clusters while the
dwarf-irregular population is the more numerous in poor groups and
in the field. However, our own Local Group seems to be an exception,
with a larger number of dwarf spheroidals than irregular galaxies.

The main goal of our project is to identify the population of
low-surface-brightness (LSB) dwarf galaxies in nearby groups to M$_V =
-10$ mag and determine the luminosity function in each of
the groups. The selected sample
is formed by groups with different environments, poor groups like
Dorado and NGC 6868 and rich groups like Hydra I and IC 4765 with
galaxy populations of different sizes and morphological mixtures. The
program was started with the observation of the Dorado group \cite{car01}, at
$cs\sim1200$ km/s, and  continued with the observation of other
seven nearby groups with  $1000<cz<4500$ km/s. The aim of this paper is
to present preliminary  results of the study of the dwarf galaxy
population in three of the groups observed, HCG42, IC4765 and NGC6868.

\section{The sample, data acquisition and photometry}

\begin{itemize} 
\item[$\bullet$] HCG42: This compact group is centered around
the giant elliptical galaxy NGC3091 (radio source) and
contains other four bright member galaxies.  

\item[$\bullet$] IC4765: This is a rich southern group  fairly close to
the galactic plane ($b\sim-23$). It is dominated by the central D galaxy
IC4765  and by the barred galaxy IC4769 (Seyfert 2) that lies 12 arcmin
away to the north. It was defined
as an interacting pair: a spiral-poor group centered around IC4765 and
a spiral-rich group centered on IC4769 \cite{qui75}.

\item[$\bullet$] NGC6868: is a loose structure dominated
by the giant elliptical galaxy NGC6868 (type E2). The group 
covers an area of $2\times2$ deg$^{2}$ on the sky with an elongated 
form in the NE-SW direction.  
\end{itemize}

Table 1 shows some relevant observed properties of the groups.

The groups were observed in February and March 1998, with the 1.3m Warsaw
telescope (Las Campanas  Observatory, Chile). The standard Johnson V and
Cousins I filters were used to obtain the images.  All data were taken
under photometric conditions except for one of the fields in NGC6868,
with a seeing ranging from 1'' to 1.3''.  The areas covered in each
group are 1425, 388 and 394 arcmin$^{2}$ for HCG42, IC4765  and
NGC6868 respectively.  For each
group, except for NGC6868, two or three images just outside the groups 
 were taken as control fields.
The targets were observed
with a small  overlapping  region between the fields in order to check
the photometry and estimate the  photometric errors. The images were
bias/overscan-subtracted, trimmed and flat fielded  using standard
procedures. The zero point calibrations were obtained using standard
stars from Landolt \cite{lan92}.

The detection, photometry and classification of the objects were
performed using the Source Extractor (SExtractor) software 
program \cite{ber96}. Before running SExtractor, we removed all bright
galaxies, saturated stars and diffuse light from the fields. After sky
subtraction, all objects with a threshold $\ge$ 1.1$\sigma$ in V ($25.7$
mag/arcsec$^2$) and $\ge$ 1$\sigma$ in I ($\sim 24.4$ mag/arcsec$^2$)
above the sky  level and with a minimum area of 10 pix$^{2}$  ($\sim1.8$
arcsec$^{2}$) were found and extracted.

Photometry of all selected objects in both filters was done using
elliptical apertures. Total magnitudes were computed using the Kron's
``first moment''  algorithm $r_{1}=\sum rI(r)/\sum I(r)$.  The catalogs
in both filters were  matched in order to obtain color information for
all objects. Colors were determined by measuring magnitudes in a circular
aperture of 3'' of diameter, in both filters.

\section{Preliminary analysis and results}

\bigskip

\begin{table}[t]
\caption{Some relevant properties of the observed groups}
\vspace{0.4cm}
\begin{center}
\begin{tabular}{lccrrcrcl}  \hline
 Group  &  $\alpha(2000)$ & $\delta(2000)$ & $V_{\odot}$ & $\sigma$ & $M_{V}$ & $N_{mem}^{(a)}$ & $A_{V}$ & Other names \\  \hline
HCG 42  & 10 00 13.0 & -19 38 29 & 3828 \cite{zab98} & 211 & 0.42 \cite{zab98} &  22 & 0.13 & \\
IC 4765 & 18 47 15.0 & -63 21 59 & 4570 \cite{mal92} & 551 & 2.97 \cite{gir98} & 118 & 0.30 & S805, DC184-630\\
NGC 6868& 20 09 43.0 & -48 17 08 & 2762 \cite{ram96} & 213 &              &   5 & 0.16 & Telescopium\\ \hline
\end{tabular}
\end{center}
{\footnotesize{(a)  number of bright galaxy members with known redshifts from the literature.}}
\end{table}

The galaxies were selected using the classification given by SExtractor,
i.e. objects with stellarity index $\le0.35$ (this index varies from
0: galaxies to 1: stars). This classification was checked by visual
inspection of a subsample of galaxies and  
is adequate down to $V\sim22.5$ for objects with stellarity index
$\le0.35$. For the brightest galaxies in each group, we calculated
the total magnitudes and colors via surface photometry using the
program ELLIPSE in the STSDAS package inside IRAF. In order to check
the photometry and estimate the errors, we have compared the total
magnitudes of all objects (galaxies and stars) in the overlapping
regions of adjacent fields. The results for the three groups show an average
difference of $0.016\pm0.020$ mag  for $V\le20$ and $\Delta V < 0.3$ for
objects close to the completeness limit ($V\le22.5$). Figure 1 shows the
color-magnitude diagram for the galaxies in the groups HCG42 and IC4765
(left and right panels respectively) and for the galaxies in the control
fields (plots on the right in both panels).

\bigskip

\begin{figure}
\psfig{figure=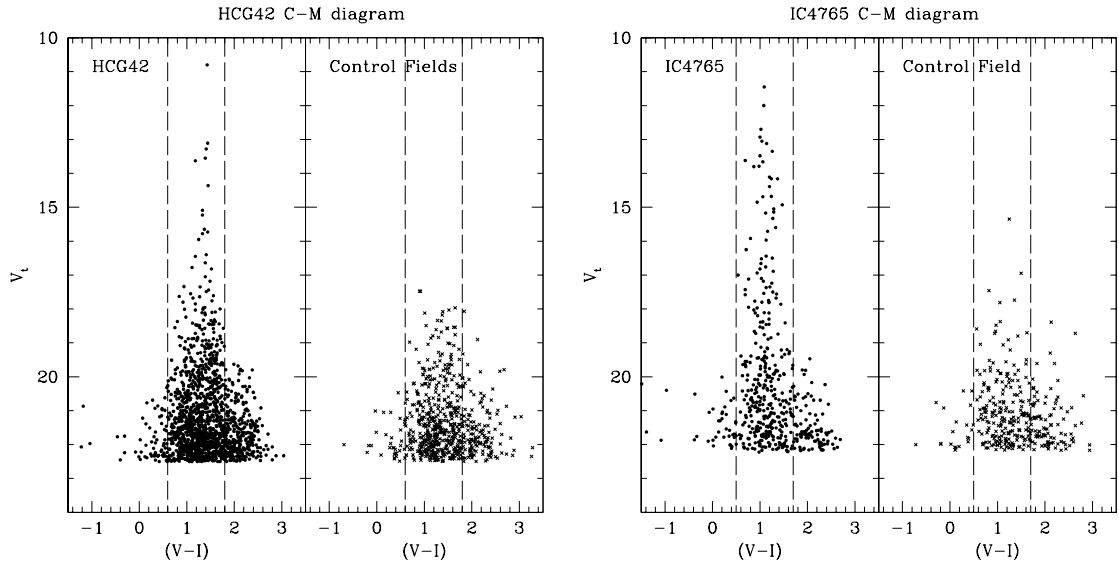,width=0.95\textwidth}
\caption{ Color-magnitude diagrams for the HCG42 (left panel) and IC4765 (right panel) 
groups and for the control fields down to a limiting magnitude of $V=22.5$. Colors and
magnitudes are corrected for interstellar extinction. The dashed lines show the
region within which the LSB are expected to lie ($0.5\le(V-I)\le1.7$).
\label{fig:radish}}
\end{figure}

The selection of the LSB dwarf galaxies was done using the color
information and the parameters given by the exponential profile fit
(central  surface brightness, the scalelength and the limiting diameter).
Fig. 1 suggests that the dwarf galaxy population may be populating
the region between $0.5\le(V-I)\le1.7$ mag (in general, the LSB dwarf
galaxy population  shows a median color of $V-I=1.0$ with a dispersion of
$\pm0.6$ mag \cite{mcg94,one97}.  As we can
see, this galaxy population is absent in the control field.  Using the
area within the broken lines in Fig. 1 to restrict our sample, we used
the structural parameters of the galaxies to select the LSB dwarf galaxy
candidates in the groups.

The radial profiles of the dwarf galaxies fainter than
$M_{V}\sim-17$ are in general well fit by an exponential
profile of the form:

\begin{equation}
\mu(r) = \mu_{0} + 1.086 (r/h)
\label{equ3}
\end{equation}

\noindent where $\mu_{0}$ is the extrapolated central surface 
brightness and $h$ is the scalelength obtained by the fit. 

The fit to a pure exponential gave the scalelength of the
dwarf galaxy candidates and was extrapolated to obtain the 
central surface brightness. These two parameters are used as a primary
cut on the sample. LSB dwarf galaxy candidates with $\mu_{0}\ge22$ V
mag/arcsec$^{2}$ and scalelength  $h>1.3$'' were selected.  The surface
brightness cut is similar to that used in the selection of 
LSB galaxies in Virgo\cite{imp88}. A second cut was done  selecting galaxies
with a limiting angular diameter larger than 4'' (HCG42), 3'' (IC4765)
and 5'' (NGC68678)  at a given isophotal level. These limiting diameters
are similar to those used to select the LSB dwarf galaxies in the Dorado
group ($0.83$ h$_{75}^{-1}$ kpc, where $h_{75}$ is the ratio of the
Hubble constant used to $H_{0}=75$ km/s/Mpc).
The limiting diameter can be expressed  as follows

\begin{equation}
\theta_{lim}=0.735 (\mu_{lim} - \mu_{0}) 10^{0.2(\mu_{0}-m_{tot})}
\label{equ5}
\end{equation}

\noindent where $\mu_{lim}$ is the surface brightness at the limiting 
isophote and $m_{tot}$ is the total magnitude of the object. 

\bigskip

\begin{figure}
\psfig{figure=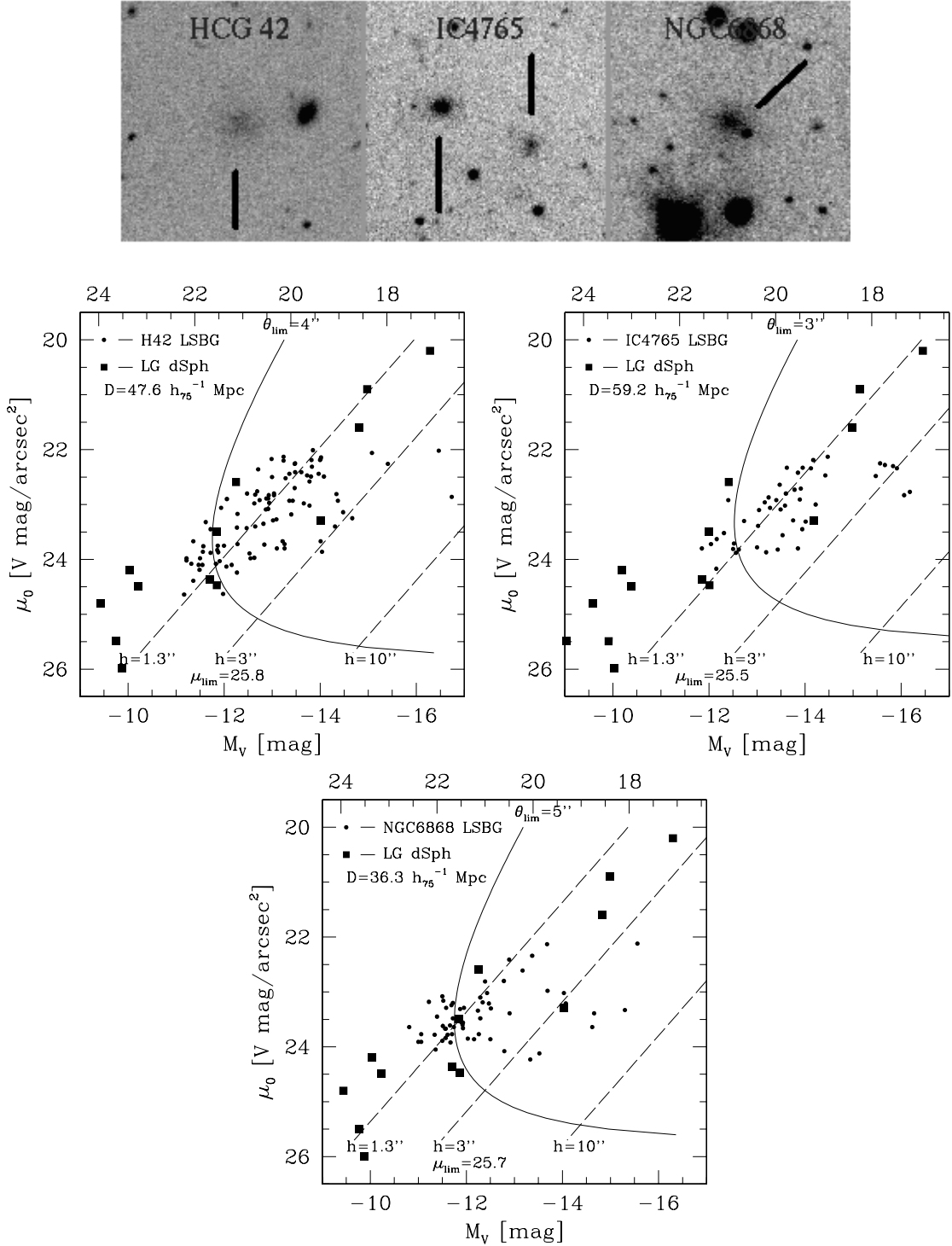,width=0.95\textwidth}
\caption{Upper panel:
a mosaic of selected LSB dwarf galaxies detected in the groups (each box has
a size of 1$\times$1 arcmin$^2$); Lower panel: extrapolated central
surface brightness versus total magnitude for LSB dwarf galaxies
detected in the HCG42, IC4765 and NGC6868  groups respectively (small
solid circles). For comparison, we show  the Local Group dSphs  galaxies
(solid squares) redshifted to the distances of the groups (46.7, 59.2  a
36.3 h$_{75}^{-1}$ Mpc for HCG42, IC4765 and NGC6868 respectively.)
\label{fig:radish}}
\end{figure}

Figure 2 shows the $\mu_{0} - V_{tot}$ plane for the LSB dwarf galaxies
detected in the HCG42, IC4765 and NGC6868 groups  after the cuts described
above (small solid circles).  For comparison, we  plotted in both diagrams
the Local Group dSph  galaxies (solid squares) redshifted to the distance
of the groups. From the figure we can see that only the brightest dSphs
of the Local Group could be detected (like Fornax and Sculptor) at the
distances of the groups (47.6, 59.2 and 36.3 $h_{75}^{-1}$ Mpc for HCG42,
IC4765 and NGC6868 respectively).

In order to check  how many galaxies we lost using the selection criteria
applied above, we performed a series of add-galaxy experiments. The
observed frames were used for these experiments to insure the proper
accounting of cosmetic defects, crowding, light gradients, noise
and seeing. For each image we simulated galaxies with exponential
profiles with scalelengths and central  surface brightnesses
typical of low-surface brightness  galaxies at the
redshift of the groups. Each artificial galaxy was  convolved with a Gaussian point
spread function constructed from bright stars in the  frame. Then,
for each bin of central surface brightness ($22\le\mu_{0}\le25$) we
generated 400 randomly distributed disk galaxies in groups of 20 (5 for
the brightest  bins)  with magnitudes  between 16.0 -- 23.0 (in bins of
0.5 mag). Then we ran the  SExtractor program with the same detection
parameters previously used for the real-galaxy detection.  After that,
the output catalogues were matched with the input ones to detect the
simulated galaxies with stellarity index $\le0.35$. The extrapolated
central surface brightnesses, the angular scalelengths and the limiting
diameters were then calculated and compared with the input parameters.

The results of the simulations for the HCG42 group are  shown in Figure 3.
The left panel shows that the completeness fraction is, on average,
$\sim 75$\% for galaxies with central surface brightness between
$22-24$ mag/arcsec$^{2}$ and brighter than $V=20$ mag. The completeness
fraction for galaxies with $22\le\mu_{0}\le23$ decreases very quickly at
fainter magnitudes. This could be an effect of a bad classification
of the Sextractor program  where the compact, high-surface brightness
objects are easily confused with stars and missed from the catalogue
(if these exist).  However, the numbers of
these objects  may be small.

For galaxies with a central surface brightness fainter than 24
V mag/arcsec$^{2}$, the completeness fraction is 60\% and lower.
The right panel in Fig. 3  shows the differences between the input and
output central surface brightnesses (in magnitudes)
 and scale factors (in percentage) of the
simulated galaxies.  To $V\sim19$,  the mean difference is about 30\% in
the scale factor, and 0.5 in $\mu_{0}$. These results are in
agreement with those obtained for the Dorado group \cite{car01}

\bigskip

\begin{figure}
\psfig{figure=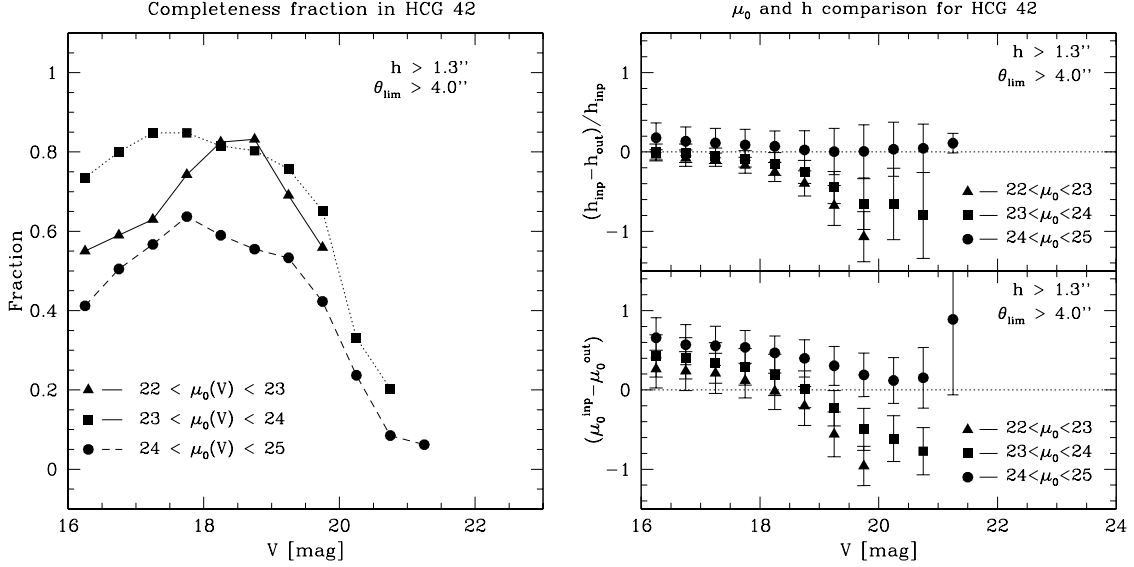,height=3.0in,width=0.95\textwidth}
\caption{
Completeness fraction as a
function of the total magnitude. Right panel: Comparison
between the input and output scalelengths (in percentage) and central surface
brightnesses for simulated galaxies.
\label{fig:radish}}
\end{figure}

\section{Summary}

More than a hundred LSB dwarf galaxy candidates with $\mu_{0}\ge22$
V mag/arcsec$^{2}$, $h\ge1.3$'', $\theta_{lim}\ge3$'' and with
$0.5\le(V-I)\le1.7$ were found in the groups HCG42, IC4765 and
NGC6868. Our add-galaxy experiment showed that the completeness fraction
was, on average, about $\sim 75$\% for galaxies with $22\le\mu_{0}\le24$,
$V\le20$ and $\sim 50$\% for lower surface brightness galaxies. After
correction by completeness and background galaxies, we  found an excess of
LSB dwarf galaxies in the fields of the groups that must be group members.
Further work will add the data of all groups
together to determine the combined luminosity function.  In addition, we
plan to study the spatial distributions, colors, types and sizes of the
dwarf galaxies as a function of the richness of their environments.

\section*{Acknowledgments}
The authors are grateful to the Director of the 1.3 Warsaw telescope for
generous allocation of telescope time and to the local staff for help with
the observations.  ERC and CMdO acknowledge support for this work provided
by FAPESP project Nrs. 96/04246-7 (ERC PhD fellowship) and 00/04399-5.

\end{document}